\documentclass[twocolumn,aps,prl,groupedaddress]{revtex4}
\usepackage{graphicx}
\usepackage{color}
\usepackage{amsmath}
\usepackage{marvosym}
\usepackage{bm}

\newcommand{\be}{\begin{equation}}
\newcommand{\ee}{\end{equation}}
\newcommand{\bea}{\begin{eqnarray}}
\newcommand{\eea}{\end{eqnarray}}

\newcommand{\la}{\langle}
\newcommand{\ra}{\rangle}

\newcommand{\ket}[1]{| #1\ra}
\newcommand{\Bra}[1]{\la #1|}
\newcommand{\Ket}[1]{| #1\ra}

\renewcommand{\vec}[1]{{\bf #1}}
\renewcommand{\epsilon}{\varepsilon}

\begin{document}
\title{Berryogenesis: self-induced Berry flux and spontaneous\\ non-equilibrium magnetism}
\author{Mark S. Rudner$^1$} 
\email{rudner@nbi.ku.dk}
\author{Justin C. W. Song$^{2,3}$}
\email{justinsong@ntu.edu.sg}
\affiliation{$^1$ Center for Quantum Devices and Niels Bohr International Academy, Niels Bohr Institute, University of Copenhagen, 2100 Copenhagen, Denmark}
\affiliation{$^2$Division of Physics and Applied Physics, Nanyang Technological University, Singapore 637371}
\affiliation{$^3$Institute of High Performance Computing, Agency for Science, Technology, and Research, Singapore 138632}

\begin{abstract}

Spontaneous symmetry breaking is central to 
the description of interacting phases of matter. Here we reveal a new mechanism through which a driven interacting system subject to a time-reversal symmetric driving field can spontaneously magnetize. We show that the strong internal ac fields of a metal driven close to its plasmon resonance may enable {\it Berryogenesis}: the spontaneous generation of a self-induced Bloch band Berry flux. The self-induced Berry flux supports and is sustained by a circulating plasmonic motion, which may arise even for a {\it linearly polarized} driving field. This non-equilibrium phase transition occurs above a critical driving amplitude, and may be of either continuous or discontinuous type. Berryogenesis relies on feedback due to 
interband coherences induced by  internal fields, and may readily occur in a wide variety of multiband systems. We anticipate that graphene devices, in particular, provide a natural platform to achieve Berryogenesis and plasmon-mediated spontaneous non-equilibrium magnetization in present-day devices.

\end{abstract} 

\maketitle

When a physical system is governed by statistical or dynamical equations possessing certain symmetries, 
its stationary states can be classified into phases according to which of those symmetries are preserved, and 
which are broken~\cite{Chaikin}.
Transitions between these different phases are characterized by spontaneous symmetry breaking -- a concept which is fundamental to the description of phases of matter in equilibrium thermodynamics. Importantly, this same concept can be applied as well to the dynamics of non-equilibrium many-body systems, where novel dynamical phases and phenomena may arise~\cite{Haken1975}. 

Time-dependent driving by laser or microwave fields has recently emerged as a powerful tool for dynamically controlling the non-equilibrium properties of quantum matter~\cite{Wang2007,Sie2016,Basov2017,Wang2013,CavalleriA,CavalleriB,CavalleriC}. Periodic driving, in particular, provides means to modify the band structures of materials on demand, and to alter the corresponding internal structure of their electronic wave functions~\cite{Oka2009,Kitagawa2011,KBRD2010,Lindner2011, Gu2011, Usaj2014, Katsnelson2015, CayssolReview}.  
These ``Floquet engineering'' proposals focus primarily on the regime where the modifications of the system's properties are induced directly by an external driving field. 

\begin{figure}[t]
\includegraphics[width=1.0 \columnwidth]{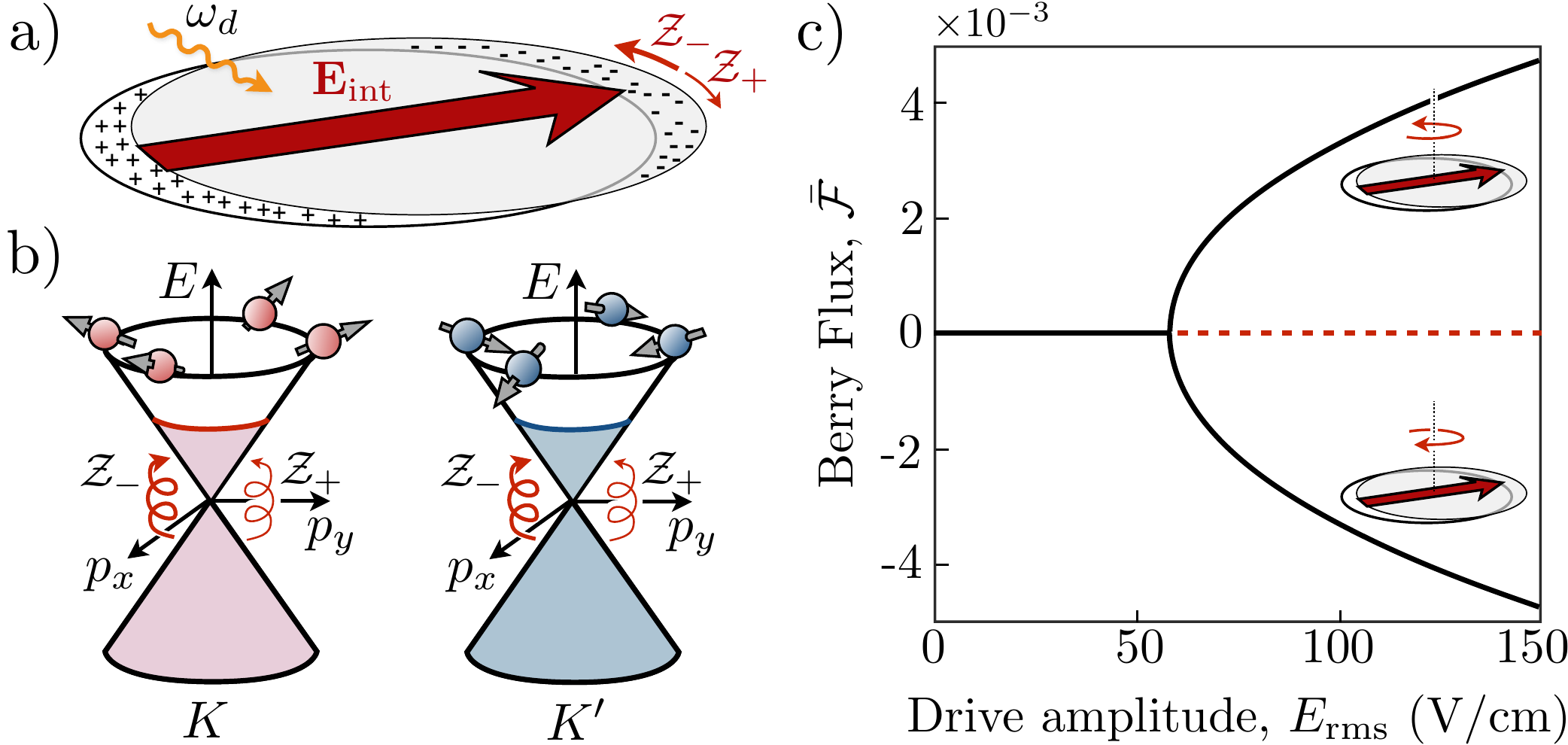}
\caption{{\it Berryogenesis}: Spontaneous generation of 
Berry flux in a time-reversal-invariant driven system.  
a) When excited, the dipole mode of a metallic disk hosts an internal electric field, $\vec{E}_{\rm int}(t)$, which oscillates at the drive angular frequency, $\omega_d$.  
The corresponding motion is decomposed into right and left circulating amplitudes, $\mathcal{Z}_+$ and $\mathcal{Z}_-$, respectively. b) An off-resonant 
ac electric field dresses 
the electronic Bloch functions, yielding a finite Berry flux through the Fermi sea. 
In this way, ac {\it internal fields} may dynamically alter the system's electronic properties, giving rise to feedback. 
c) Stability diagram, with stable (unstable) solutions of Eq.~(\ref{eq:SelfConsistent})  indicated by black solid (red dashed) lines.
For small amplitudes of a  linearly polarized drive, the response is linearly polarized ($\bar{\mathcal{F}} = 0$).
Above threshold, the linearly polarized solution becomes unstable and the system spontaneously develops a right- or left- handed circulation. Parameters: $E_F = 160$ meV, $\hbar\omega_d = 100$ meV, $\hbar \omega_0 = 100.125$ meV, $Q = 75$.
}
\label{fig:intro}
\vspace{-0.2 in}
\end{figure}

Importantly, interacting systems may also host strong {\it internal} fields when pushed out of equilibrium. 
For example, plasmons have recently gained wide attention for their ability to resonantly enhance applied electric fields by many orders of magnitude~\cite{Atwater2010,Koppens2011}.
Here we propose that interesting new phase structures may arise due to {\it feedback} in which such internal fields modify a system's electronic properties, in turn altering 
its response to the driving field.
As we show, in a metallic disk driven at a frequency close to its natural dipole resonance (Fig.~\ref{fig:intro}{\bf a}), this feedback gives rise to multistability, and, intriguingly, can cause the system to spontaneously develop a chiral circulating motion even when it is driven by a linearly polarized field (Fig.~\ref{fig:intro}{\bf c}). 

The instability toward the magnetized state 
is driven by {\it Berryogenesis}: the spontaneous generation of Bloch band Berry flux.  
To illustrate this phenomenon we study 
the nonlinear dynamics of the plasmonic dipole resonance of a metallic disk.
When excited, the resonance may produce rotating fields that modify the system's Bloch band Berry curvature (Fig.~\ref{fig:intro}{\bf b}). 
The resulting amplitude-dependent Berry flux splits the resonance for right- and left-handed polarizations, 
yielding a nonlinear feedback between field strength, field chirality, and Berry flux, that ultimately leads to spontaneous symmetry breaking. 

The threshold driving amplitudes that are needed to induce Berryogenesis 
are modest, and 
can be achieved in current high quality graphene
devices~\cite{Achim2015,Xi2018,Iranzo2018} with readily available terahertz to mid-infrared drives. 
While we use graphene 
for illustration due to its simplicity and favorability for near-term experiments, the mechanism we discuss is general and may apply to a wide variety of multiband materials. 
The ease with which large plasmon-enhanced internal electric fields can be accessed in graphene~\cite{Koppens2011,Iranzo2018} makes it a natural platform for 
plasmonic non-equilibrium spontaneous symmetry breaking. 

{\bf Berry flux generation ---} 
We now demonstrate how a nonvanishing Berry flux can be induced in a time-reversal-symmetric metal subjected to an off-resonant optical field.
As a concrete example we consider electrons in graphene at finite doping, with Fermi energy $E_F > 0$.
This setup provides a simple illustration for how light-matter interaction generates the non-trivial wave function coherences 
that lead to {\it Berryogenesis}. 

Taking $E_F$ as the natural energy scale, we express the Hamiltonian for electrons in the $K$ valley of graphene as
\be
\mathcal{H}_K = E_F\big[\tilde{\vec k} - \tilde{\vec{A}}(t) \big] \cdot \boldsymbol{\sigma},\quad \tilde{\vec{A}} (t) = \frac{ev}{cE_F} \vec{A} (t),
\label{eq:graphene}
\ee
where $v$ is graphene's Fermi velocity, $c$ is the speed of light, $-e < 0$ is the electron charge, $\boldsymbol{\sigma} = (\sigma_x, \sigma_y)$ is a vector of Pauli matrices, and $\tilde{\vec{k}} = \vec{k}/k_F$ is the normalized 2D wave vector, with $\hbar v k_F = E_F$. The vector potential $\vec{A}(t)$ describes the (time-varying) electromagnetic (EM) field.
Importantly, $\vec{A}(t)$ may arise both due to an externally applied driving field, and due to time-dependent internal fields of the system when it is driven out of equilibrium.
The Hamiltonian $\mathcal{H}_{K'}$ for electrons in the $K'$ valley is the same as that in Eq.~(\ref{eq:graphene}), with $\sigma_y\rightarrow -\sigma_y$ in $\boldsymbol{\sigma}$.

For illustration we focus on a monochromatic field, $\vec{A}(t)$, oscillating at frequency $\omega$.
Crucially, we will consider off-resonant frequencies with $\hbar\omega < 2E_F$ such  that absorptive transitions are prevented by Pauli blocking. 
In this situation, the main action of the 
time-varying $\vec A(t)$ in Eq.~(\ref{eq:graphene}) is to dress/hybridize the wave functions in the conduction and valence bands of $\mathcal{H}_0$~\cite{Wang2007}. Below it will be convenient to work in the basis of circularly polarized fields, $\vec{A}(t)= \frac12 (\vec A_L + \vec A_R) e^{-i\omega t} \,+\,c.c.$, where the left- and right-handed components are given by $\vec{A}_L = A_L(\hat{\vec{x}} + i \hat{\vec{y}})/\sqrt{2}$ and $\vec{A}_R = A_R(\hat{\vec{x}} - i\hat{\vec{y}})/\sqrt{2}$.

We analyze the induced Berry flux by studying the Floquet-Bloch band structure arising from Eq.~(\ref{eq:graphene}) with the time-periodic field $\vec{A}(t)$.
For each $\vec{k}$ we find two Floquet state solutions to the time-dependent Schr\"{o}dinger equation~\cite{Shirley65, Sambe73}: $\Ket{\psi_{\vec{k}\alpha}(t)} = e^{-i\epsilon_{\vec{k}\alpha}t}\Ket{\Phi_{\vec{k}\alpha}(t)}$, where $\alpha = \pm$ is the Floquet band index, $\epsilon_{\vec{k}\alpha}$ is the corresponding quasienergy, and $\Ket{\Phi_{\vec{k}\alpha}(t)}$ is a periodic function of time, with period $T = 2\pi/\omega$. Due to the off-resonant nature of the ac-field, states at the Fermi surface are only weakly modified; 
we assume that their populations map smoothly onto the Floquet states.

The periodic part of the Floquet eigenstate, $\Ket{\Phi_{\vec{k}\alpha}(t)}$, typically involves frequency components at many harmonics of the drive frequency $\omega$.
As a result, the Berry connection $\mathcal{A}_{\vec{k}\alpha}(t) = \Bra{\Phi_{\vec{k}\alpha}(t)}i\nabla_{\vec{k}}\Ket{\Phi_{\vec{k}\alpha}(t)}$, and ultimately the net Berry flux in the (dressed) conduction band~\cite{Haldane2004}, $\mathcal{F}(t) = \oint d\vec{k} \cdot \mathcal{A}_{\vec{k}+}(t)$, will be periodic functions of time. 
(Here the integral is taken around the Fermi surface.)
In the analytical treatment below we will focus on the steady-state (dc) part of the induced Berry flux, $\bar{\mathcal{F}}$, which captures the net breaking of time-reversal symmetry (TRS) and provides the driving force for {\it Berryogenesis}. Through fully self-consistent numerical simulations we will show that including the ac part of $\mathcal{F}$ does not significantly alter our conclusions.

Due to TRS, the time-averaged Berry flux $\bar{\mathcal{F}}$ 
must vanish for a linearly polarized field, i.e., for $|{A}_L| = |{A}_R|$.
However, 
a chiral field with $|{A}_L| \neq |{A}_R|$ breaks TRS. In this case, the field-induced band hybridization is manifested as a canting of the electronic pseudospins 
(Fig.~\ref{fig:intro}{\bf b}), which yields a finite Berry flux, $\mathcal{F} \neq 0$~\cite{Oka2009,Kitagawa2011,Gu2011, Usaj2014, Katsnelson2015}.

\begin{figure}[t!]
\includegraphics[width=1.0 \columnwidth]{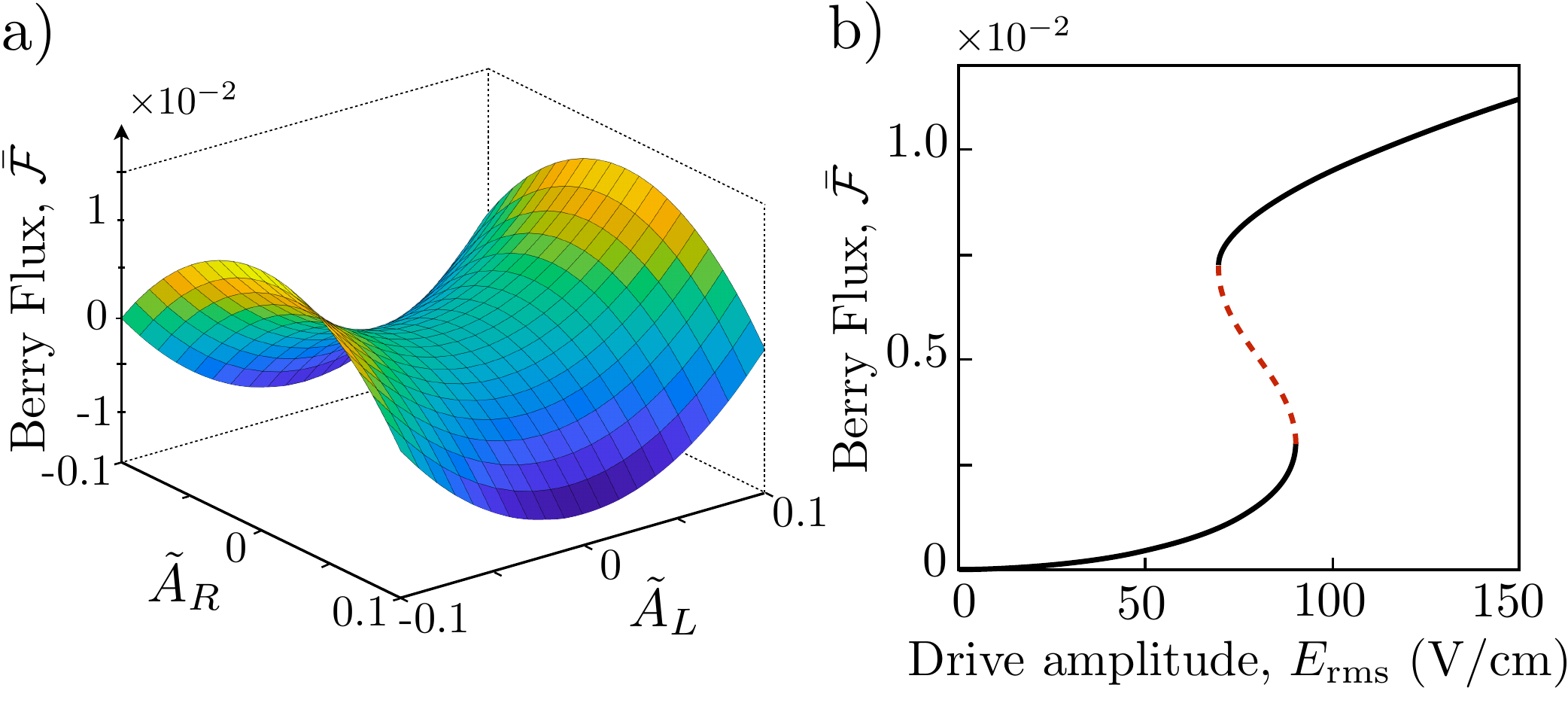}
\caption{Berry flux generation and non-linear plasmon dynamics. a) Time-averaged Berry flux induced by a harmonic drive with angular frequency $\omega_d$ and (dimensionless) left- and right-circularly polarization amplitudes $\tilde{A}_L$ and $\tilde{A}_R$, respectively [see Eq.~(\ref{eq:graphene})].  In the off-resonant regime, the induced flux is approximately given by $\tilde{F} \approx \beta (|\tilde{A}_L|^2 - |\tilde{A}_R|^2)$, where $\beta \sim \mathcal{O}(1)$ depends on $\hbar\omega_d/E_F$. The surface shown is for $\hbar\omega/E_F = 0.625$. b) Steady state Berry flux for a graphene disk driven by a circularly polarized driving field. Stable and unstable steady states are indicated by black solid and red dashed lines, respectively. Parameters: $E_F = 160$ meV, $\hbar\omega_d = 100$ meV, $\hbar\omega_0 = 100.75$ meV, $Q = 100$.}
\label{fig:CircularPolarization}
\end{figure}

In Fig.~\ref{fig:CircularPolarization}{\bf a} we show the dc Berry flux induced by an off-resonant field, as a function of the (scaled)  left- and right-hand circulating field amplitudes $\tilde{A}_L$ and $\tilde{A}_R$.
In this regime, $\bar{\mathcal{F}}$ exhibits a characteristic saddle-like shape $\bar{\mathcal{F}}  \approx \beta ( |\tilde{A}_L|^2 - |\tilde{A}_R|^2)$, as may be anticipated by second-order perturbation theory.
Here $\beta$ is an $\mathcal{O}(1)$ dimensionless prefactor; for the parameters in graphene used in Fig.~\ref{fig:CircularPolarization}{\bf a} we find $\beta \approx 2.3$.
These considerations show that Berry flux generation both provides a sensitive detector for TRS breaking in electronic systems, and as we will see below, naturally couples with plasmonic dynamics.

{\bf Non-linear plasmon dynamics ---} Above we showed that a circulating electric field $\vec{E}(t) = -\tfrac{1}{c}\frac{\partial}{\partial t} \vec{A}(t)$ induces a non-vanishing 
net Berry flux in a time-reversal invariant metal. 
Crucially, when a plasmonic mode is excited, $\vec A(t)$ in Eq.~(\ref{eq:graphene}) includes an {\it internal field} contribution due to the oscillating charge density in the system.
Near a resonance, $\vec{A}(t)$ may easily be dominated by these plasmonic/internal fields~\cite{Atwater2010,Koppens2011}.
As we now illustrate, the plasmon resonance itself is sensitive to the development of Berry flux; this characteristic interdependence (feedback) underlies {\it Berryogenesis}, see next section.  

To illustrate Berry flux generation via plasmonic fields, we consider the dipolar mode of a circular electronic disk (Fig.~\ref{fig:intro}{\bf a}).
We describe the plasmon dynamics using center of mass (COM) coordinates for position, $ \{ \vec r (t) \} = \{x (t), y (t)\}$, and momentum, $\{ \vec p(t)\} = \{p_x (t), p_y (t)\}$. 
Here $\{ \cdot \}$ denotes 
an average over the electronic distribution function. The resulting plasmon equations of motion (derived from the kinetic equation~\cite{Son2012, SonSpivak2013, Song2016}) are given by: 
\begin{align}
\frac{d\{\vec{r}\}}{dt} & =  \frac{\{\vec p\}}{m} - \frac{{\mathcal{F}[\vec{E}_{\rm tot}(t)]}}{\hbar n_0} \hat{\vec z} \times e\vec E_{\rm tot}(t),  
\nonumber \\
\frac{d\{\vec{p}\}}{dt} & = - m\omega_0^2 \{\vec{r}\} - \gamma \{\vec p\} - e\vec E_{\rm drive} (t), 
\label{eq:dynamicsfull}
\end{align}
where $\omega_0$ is the angular frequency of the bare plasmon resonance of the disk (i.e., for ${\mathcal{F}} = 0$), $m$ is the plasmon mass, $n_0$ is the electron density, and $\gamma$ is the damping rate of the plasmon mode~\footnote{We include $\gamma$ in the equation of motion to account for the plasmon damping that inevitably arises in real devices.  However, dissipation is {\it not} essential for multistability or spontaneous symmetry breaking in this system.}.
The fields $\vec E_{\rm tot} (t)$ and $\vec E_{\rm drive}(t)$ are the total and the (monochromatic) driving electric fields, respectively. 
Importantly, for ${\mathcal{F}} \neq 0$, the cross product in the first line of Eq.~(\ref{eq:dynamicsfull}) couples the modes with linear polarizations along $\hat{\vec{x}}$ and $\hat{\vec{y}}$.
The nonlinearity of the system arises due to the dependence of 
$\vec{E}_{\rm tot}$ on the plasmonic motion, $\{\vec{r}(t)\}$ (see below).

Within our mean-field approach, that focuses on the COM motion of the plasmon dipole, any spatial dependence of the internal and external fields is integrated out to obtain the effective electric field acting on the COM: $e\vec E_{\rm tot} (t) = e\vec E_{\rm drive} (t) +  m\omega_0^2 \{\vec{r}\}$, where $m\omega_0^2 \{\vec{r}\}$ is the restoring force acting on the COM. 
Close to resonance, the plasmonic internal field is enhanced by a factor $\omega_0/\gamma = 2Q$ relative to the driving field, where $Q$ is the quality factor of the resonance.
For large $Q$~\cite{Achim2015, Xi2018, Iranzo2018}, the total electric field may thus be well approximated by $e\vec E_{\rm tot} (t) \approx  m\omega_0^2 \{\vec{r}(t)\}$; for simplicity, we  make this replacement in the analysis below.

Our approach [Eq.~(\ref{eq:dynamicsfull})] is designed to capture the time-periodic {\it steady-state} motion of the system, where the Berry flux ${\mathcal{F}}$ is determined self-consistently from the (periodic) oscillating external and internal fields present in the steady-state. Transient dynamics, including fluctuations in the vicinity of the phase transition~\cite{DykmanKrivoglaz, DykmanBook}, are beyond the scope of this work. Noting that the time-averaged part of the Berry flux plays the most essential role in altering the character of the plasmon dynamics, below we replace $\mathcal{F}$ by its time-averaged value, $\bar{\mathcal{F}}$, in Eq.~(\ref{eq:dynamicsfull}).
This approximation preserves the crucial nonlinearity associated with the dependence of $\bar{\mathcal{F}}$ on 
$\{\vec{r}(t)\}$, and enables a detailed analytical treatment.
We support these results with fully self-consistent numerical simulations, which incorporate the time dependence of $\mathcal{F}(t)$. 

We now  solve for the steady-state plasmonic oscillations described by Eq.~(\ref{eq:dynamicsfull}) with $\mathcal{F}$ replaced by $\bar{\mathcal{F}}$. 
We work in a complex representation with $\vec{E}_{\rm drive}(t) = \vec{E}_{\rm drive}^{(0)}\,e^{-i\omega_dt}$, where $\omega_d$ is the angular frequency of the drive. 
Importantly, keeping only the dc part of the (self-generated) Berry flux yields monochromatic solutions $\{\vec{r}(t)\} = \vec{r}^{(0)}e^{-i\omega_d t}$, $\{\vec{p}(t)\} = \vec{p}^{(0)}e^{-i\omega_d t}$; the physical solutions are given by the real parts of these quantities.

Feedback due to self-generated Berry flux arises via the saddle-like dependence of $\bar{\mathcal{F}}$ on $\tilde{A}_{L}$ and $\tilde{A}_R$ (see Fig.~\ref{fig:CircularPolarization}{\bf a}).
To capture this interplay, we decompose the steady state motion into right (+) and left (-) circulating components:
$\mathcal{Z}^{(0)}_\pm = \frac{1}{\sqrt{2}}[x^{(0)} \pm i y^{(0)}]$.
We identify the internal field contribution to $\vec{A}(t)$ in Eq.~(\ref{eq:graphene}) by using $\vec E_{\rm tot} (t)= - \frac{1}{c} \frac{\partial}{\partial t} \vec{A}(t)$, together with the replacement $e\vec E_{\rm tot} (t) \approx  m\omega_0^2 \{\vec{r}\}$.
In this way we obtain the self-generated (dc) Berry flux:
\be
\bar{\mathcal{F}} = f(|\mathcal{Z}_-^{(0)}|^2/l^2, |\mathcal{Z}_+^{(0)}|^2/l^2),  \quad l^{-1} = \frac{vm\omega_0^2}{ E_F \omega_d},
\label{eq:FluxGen}
\ee
where $f$ is the (dimensionless) saddle function of Fig.~\ref{fig:CircularPolarization}{\bf a} and 
$l$ defines an intrinsic length scale of the system~\cite{SI}. 

Transforming Eq.~(\ref{eq:dynamicsfull}) to the circular polarization basis, and defining $\mathcal{E}^{\pm}_{\rm drive} (t) = \frac{1}{\sqrt{2}}[E_{\rm drive}^x (t) \pm iE_{\rm drive}^y (t)] = \mathcal{E}_\pm^{(0)} e^{-i\omega_d t}$, we find that the amplitudes $\mathcal{Z}_\pm^{(0)}$ are given by: 
\be
\mathcal{Z}_\pm^{(0)} = \frac{-e\mathcal{E}_\pm^{(0)}/m}{[-\omega_d^2 + \omega_0^2 \pm \kappa\bar{\mathcal{F}}\omega_d] - i\gamma [\omega_d \mp \kappa\bar{\mathcal{F}}]}, 
\label{eq:PlasmonAmplitude}
\ee
where $\kappa = m \omega_0^2/(\hbar n_0)$. 
Equation (\ref{eq:PlasmonAmplitude}) demonstrates how a dc Berry flux $\bar{\mathcal{F}}$ modifies the disk's dipole resonance~\cite{Song2016}.

Importantly, because $\bar{\mathcal{F}}$ in Eq.~(\ref{eq:PlasmonAmplitude}) depends on $\mathcal{Z}_\pm^{(0)}$ via Eq.~(\ref{eq:FluxGen}), the system may exhibit multistability. 
For demonstration, we first consider the simple case of a circularly polarized drive, $\mathcal{E}^{(0)}_+ = 0$, $\mathcal{E}^{(0)}_- = E_{\rm rms}$.
Here, the steady state motion captured by Eq.~(\ref{eq:PlasmonAmplitude}) is itself circularly polarized: there is no mixing between left and right hand polarizations.
In Fig.~\ref{fig:CircularPolarization}{\bf b} we show the corresponding solutions of Eq.~(\ref{eq:PlasmonAmplitude}), using the parametrized form $\bar{\mathcal{F}} = \beta (|\tilde{A}_L|^2  - |\tilde{A}_R|^2) = \beta l^{-2}( |\mathcal{Z}_-^{(0)}|^2  - |\mathcal{Z}_+^{(0)}|^2 )$.
We track the bistability via the induced (dc) Berry flux, $\bar{\mathcal{F}}$, as it provides a sensitive measure of the amplitude of circular motion. 
Note that for a circularly-polarized driving field in Eq.~(\ref{eq:graphene}), 
the induced Berry flux $\mathcal{F}$ is in fact time-independent: $\mathcal{F}(t) = \bar{\mathcal{F}}$.

In the absence of driving, the graphene disk possesses zero Berry flux, $\mathcal{F} = 0$.
As shown in Fig.~\ref{fig:CircularPolarization}{\bf b}, as the amplitude of the circularly polarized drive is increased from zero, a finite Berry flux is generated.
Strikingly, when the drive amplitude is strong enough, the induced Berry flux exhibits {\it bistability}: two distinct steady-state Berry fluxes [corresponding to two stable steady-state amplitudes for $\mathcal{Z}_-^{(0)}$ in Eq.~(\ref{eq:PlasmonAmplitude})] may arise for the same drive amplitude. 
For even stronger driving, only the solution with a large self-generated contribution to the Berry flux remains.

Bistability arises from the fact that the (self-induced) Berry flux splits and shifts the plasmon resonances of the disk. 
Consider a weak external drive, with frequency $\omega_d$ slightly red-detuned from the bare resonance $\omega_0$.
The drive induces a small amplitude circular motion of the plasmon dipole, which correspondingly generates a small Berry flux, $\bar{\mathcal{F}}$.
Due to the nonvanishing $\bar{\mathcal{F}}$, the resonance shifts downward, {\it toward} the frequency of the drive~\footnote{The direction of the shift is controlled by microscopic parameters of the system; for a right- or left-handed drive the sign is the same, but for hole-doped graphene with $E_F < 0$, for example, the sign would be reversed.}.
As the resonance approaches $\omega_d$, the amplitude of response increases, thereby amplifying $\bar{\mathcal{F}}$ and bringing the drive even closer to resonance~\footnote{For a blue-detuned drive, {\it negative feedback} is obtained as the resonance is pushed away from the drive frequency. In this case, no multistability is expected.}. 
This feedback provides the mechanism for bistability, and can lead to spontaneous magnetization as we discuss next.

{\bf Berryogenesis: spontaneous magnetization ---} We now turn our attention to the main phenomenon of interest:  
the spontaneous generation of Berry flux in a time-reversal symmetric driven system.
Above we discussed bistability in the case where a circularly polarized drive explicitly broke the TRS of the system.
There, the drive seeded the development of a Berry flux, which was then strongly enhanced by the system's internal fields.
Here we consider the case of a {\it linearly polarized} drive, which preserves TRS.
Remarkably, above a threshold drive amplitude the system spontaneously picks a direction of circulation, thereby generating its own Berry flux.

We characterize the spontaneous magnetization of the system via the order parameter $\eta \equiv |\mathcal{Z}_+^{(0)}|^2 - |\mathcal{Z}_-^{(0)}|^2$.
A non-zero value of $\eta$ indicates the presence of an internal field exhibiting a net right ($\eta > 0$) or left ($\eta < 0$) handed rotation; as discussed above, such a rotating field induces a finite Berry flux. 
Once a finite $\mathcal{F}$ is present, the shifting of resonances helps to reinforce and amplify the circulating motion.
This feedback is the driving force that enables {\it Berryogenesis} in time-reversal symmetric systems.

\begin{figure}[t]
\includegraphics[width=1.0 \columnwidth]{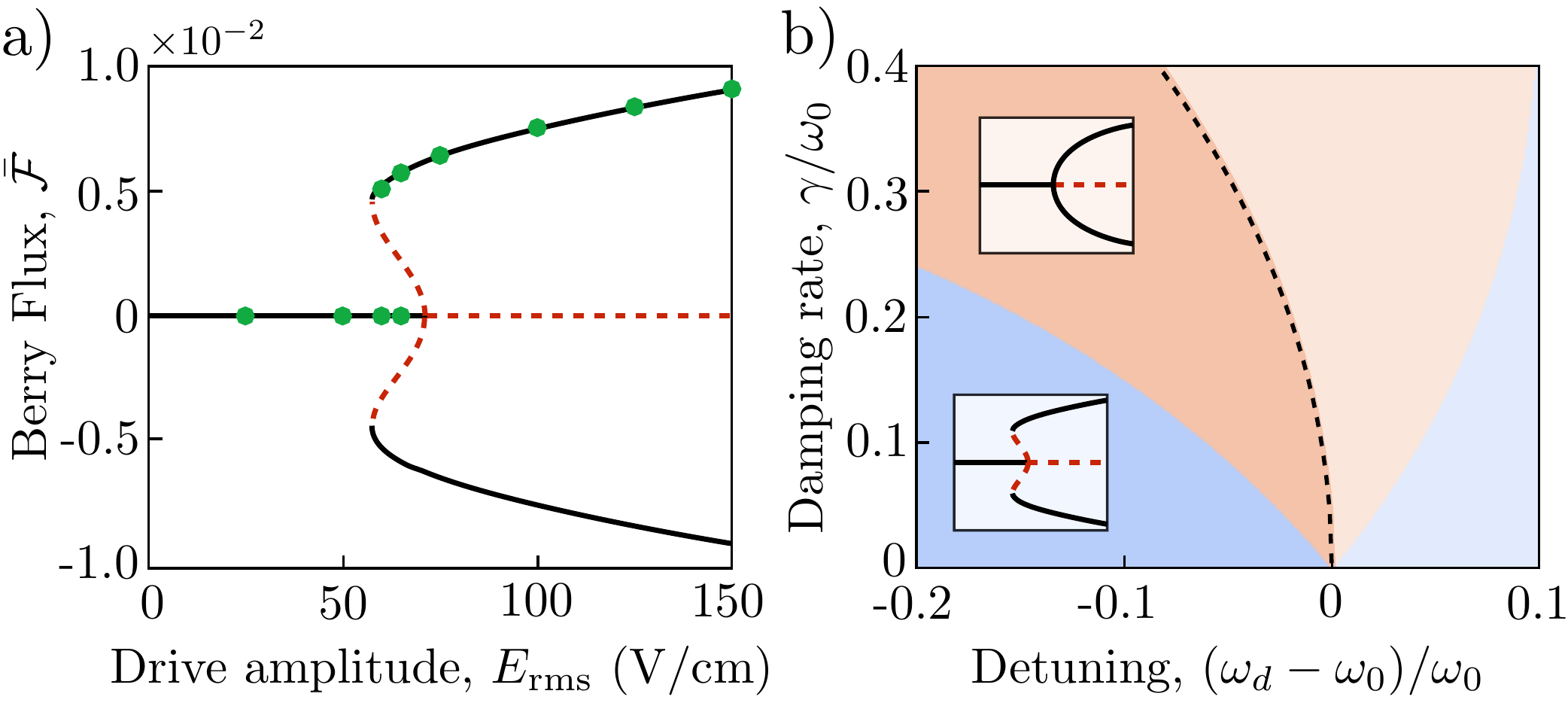}
\caption{Spontaneous magnetization in the presence of a time-reversal invariant, linearly-polarized drive.
a) Stability diagram in the discontinuous transition regime. 
Stable (unstable) solutions of Eq.~(\ref{eq:SelfConsistent}) are indicated by black solid (red dashed) lines.
Green circles indicate self-consistent solutions to Eq.~(\ref{eq:dynamicsfull}) in which the full time-dependence of $\mathcal{F}(t)$ 
is calculated from the self-induced Floquet band structure.
Parameters: $E_F = 160$ meV, $\hbar\omega_d = 100$ meV, $\hbar \omega_0 = 100.5$ meV, $Q = 100$.
b) The phase transition to the magnetized state can be either continuous (pink region) or discontinuous (blue region), depending on the detuning and damping rate.
No instability occurs in the region to the right of the dashed line.
}
\label{fig:LinearPolarization}
\end{figure} 

To illustrate spontaneous TRS breaking, we return to Eq.~(\ref{eq:PlasmonAmplitude}), with 
$\mathcal{E}^{(0)}_+ = \mathcal{E}^{(0)}_- = E_{\rm rms}$.
For the simple parametrized saddle form considered above, the dc Berry flux is a function of $\eta$ alone: $\bar{\mathcal{F}} = -\beta l^{-2} \,\eta$.
Taking the absolute value squared of both sides of Eq.~(\ref{eq:PlasmonAmplitude}) and subtracting the expressions for $|\mathcal{Z}_-^{(0)}|^2$ and $|\mathcal{Z}_+^{(0)}|^2$, we obtain an algebraic relation for the steady state magnetization:
\be
\eta\left[1 + 4{\nu} \omega_d (\omega_d^2 + \gamma^2 - \omega_0^2) \frac{|eE_{\rm rms}/m|^2}{D_+ D_-}\right]  = 0,  
\label{eq:SelfConsistent}
\ee
with $D_\pm \!=\! [\omega_0^2 - \omega_d^2 \mp {\nu} \omega_d \eta]^2 + \gamma^2 (\omega_d \pm {\nu} \eta)^2$ and ${\nu} = \kappa \beta/l^{2}$. 

Equation~(\ref{eq:SelfConsistent}) can be expressed as a fifth-order polynomial in $\eta$, and may exhibit several solutions (see Figs.~\ref{fig:intro}{\bf c} and \ref{fig:LinearPolarization}{\bf a}). 
For small amplitudes of the linearly-polarized drive, the system responds with linearly-polarized oscillations ($\eta = 0$). 
As the drive amplitude is increased, a bifurcation is encountered where the linearly polarized solution becomes unstable to fluctuations and 
the system spontaneously acquires a magnetization ($\eta \neq 0$)~\footnote{The handedness of the steady-state motion is chosen {\it spontaneously}, much as a ferromagnet spontaneously chooses a magnetic orientation when cooled below its Curie temperature.}.

We confirm the validity of the results above, which were obtained by replacing $\mathcal{F}$ by $\bar{\mathcal{F}}$ in Eq.~(\ref{eq:dynamicsfull}), by numerical solving Eq.~(\ref{eq:dynamicsfull}) including the full time dependence of $\mathcal{F}(t)$. We seek time-periodic solutions $\{\vec{r}(t)\}$ and $\{\vec{p}(t)\}$ which satisfy Eq.~(\ref{eq:dynamicsfull}), with $\mathcal{F}(t)$ calculated self-consistently from the system's Floquet band structure induced by the combination of external and internal fields, $\vec{E}_{\rm tot}(t)$ (see {\bf SI}~\cite{SI}).
As shown in Fig.~\ref{fig:LinearPolarization}{\bf a}, the time-averaged magnetization obtained from these simulations (green dots) is in good agreement with the results of our analytical treatment above (solid lines).
With the time-dependence of $\mathcal{F}(t)$ included, the solutions exhibit higher harmonic generation; within the regime studied, we find that the amplitudes of the higher harmonics are very small and do not significantly affect the behavior.

Interestingly, the type of phase transition (discontinuous vs.~continuous) is controlled by the detuning of the drive, $(\omega_d - \omega_0)/\omega_0$,  and the damping rate, $\gamma$. 
The character of the transition can be straightforwardly extracted from the $\eta$ dependence of the expression in brackets in Eq.~(\ref{eq:SelfConsistent}), see {\bf SI}~\cite{SI}.
As summarized in Fig.~\ref{fig:LinearPolarization}{\bf b}, we find that: i) spontaneous magnetization may only occur for $(\omega_d^2 + \gamma^2 - \omega_0^2) < 0$, and ii) for small detunings, discontinuous transitions are favored at low damping (high Q).
Note that spontaneous symmetry breaking persists for $\gamma = 0$. 
Inspecting Eq.~(\ref{eq:SelfConsistent}) and Fig.~\ref{fig:LinearPolarization}b for $\gamma = 0$, we see that in the absence of damping the transition occurs on the red detuned side of the resonance, and is always of discontinuous type ($B < 0$).

{\bf Discussion ---} The phenomena described above are enabled by plasmonic enhancement of applied fields, which for a high quality resonance can amplify the driving field by several orders of magnitude. 
The threshold driving amplitude for Berryogenesis is therefore controlled by the quality factor of the plasmonic device, as well as by the detuning from resonance (along with intrinsic parameters such as the Fermi energy).
For the high quality factors ($Q > 100$) recently achieved in graphene plasmonic devices~\cite{Xi2018}, we expect Berryogenesis to be achievable at moderate driving powers of order 30 W/cm$^2$ at a frequency of about 25 THz (see {\bf SI}~\cite{SI}).
 
Throughout this work we have focused on feedback arising from self-generated Berry flux.
At high excitation amplitudes, nonlinear dissipation and other sources of nonlinearity (such as self-induced anisotropy of the electronic Drude weight) may also arise.
Crucially, we work at frequencies outside the particle-hole continuum, $\hbar\omega_d < 2 E_F$, where both direct absorption from the drive (which leads to heating) and the decay of a single plasmon into a particle-hole pair is forbidden by Pauli exclusion.
To suppress the nonlinear contributions to damping and heating, it is furthermore beneficial to work at lower frequencies where two- or three-photon processes are also blocked; for the simulations in this work we used $2E_F > 3 \hbar\omega_d$, guaranteeing that the rates of these intrinsic dissipative processes are small throughout the parameter range we studied.
Furthermore, we have checked that driving-induced anisotropies are also small throughout this regime, and are not expected to significantly affect the threshold for Berryogenesis (see {\bf SI}~\cite{SI}).

Berryogenesis is a ``self-Floquet'' process through which the collective motion of an electronic system causes it to reconstruct its own band structure, yielding dramatic effects including non-equilibrium spontaneous TRS breaking. Looking ahead, we anticipate that other types of non-equilibrium phase transitions, including complex spatiotemporal dynamics in extended systems~\cite{Radzihovsky2005}, may be triggered by analogous feedback mechanisms. This work opens new prospects for exploiting the near-field compression of electromagnetic fields in metals to realize novel non-equilibrium phases of matter.

{\bf Acknowledgements---} We thank D.~Huse, M.~Kats, M.~Katsnelson, L.~Levitov, R.~Nandkishore, L.~Radzihovsky, and G.~Refael for helpful discussions. M.S.R. gratefully acknowledges the support of the European Research Council (ERC) under the European Union Horizon 2020 Research and Innovation Programme (Grant Agreement No. 678862), and the Villum Foundation. J.C.W.S. gratefully acknowledges the support of the Singapore National Research Foundation (NRF) under NRF fellowship award NRF-NRFF2016-05, and a start-up grant from the Nanyang Technological University.

\newpage
\null\thispagestyle{empty}
\newpage

\setcounter{equation}{0}
\renewcommand{\theequation}{S-\arabic{equation}}
\makeatletter
\renewcommand\@biblabel[1]{S#1.}

\section{Supplementary Information for ``Berryogenesis: self-induced Berry flux and spontaneous non-equilibrium magnetization''} 

\section{Self-induced Berry flux}

In the following we discuss how internal rotating electric fields, captured by a circulating plasmon center of mass (COM) coordinate $\mathcal{Z}^{(0)}_\pm = \frac{1}{\sqrt{2}}[x^{(0)} \pm i y^{(0)}]$, directly generates a Berry flux of the electrons.  To do this, we first recall that oscillating electric fields $\vec{E}(t) = \vec{E}^{(0)}\,e^{-i\omega_dt}$ can be expressed in terms of a time varying vector potential: $\vec{A}(t) = \vec{A}^{(0)}\,e^{-i\omega_dt}$, where 
\be
\vec{E}^{(0)} = (i\omega/c) \vec{A}^{(0)}.
\label{eq:EtoA}
\ee
As in the main text, we use the circular polarization basis to describe the vector potential fields: $\vec{A}(t)= \frac12 (\vec A_L + \vec A_R) e^{-i\omega t} \,+\,c.c.$, where the left- and right-handed components are given by $\vec{A}_L = A_L(\hat{\vec{x}} + i \hat{\vec{y}})/\sqrt{2}$ and $\vec{A}_R = A_R(\hat{\vec{x}} - i\hat{\vec{y}})/\sqrt{2}$.

Circularly polarized fields modify the system's band structure and electronic wave functions.
In this way the Fermi sea may acquire a finite 
 (dc) Berry flux: 
\be
\bar{\mathcal{F}} = f( |\tilde A_L|^2 , |\tilde A_R|^2), \quad \tilde{\vec{A}}_{L,R}= \frac{ev}{cE_F} \vec{A}_{L,R},
\label{eq:fluxsupplement}
\ee
where $f$ is the (dimensionless) saddle shown in Fig.~\ref{fig:CircularPolarization}{\bf a}. 

Crucially, the fields in Eq.~(\ref{eq:fluxsupplement}) consist of both external driving fields as well as {\it internal} plasmonic fields. Within our mean-field approach, the total electric field on the COM can be expressed $e\vec E_{\rm tot} (t) = e\vec E_{\rm drive} (t) +  m\omega_0^2 \{\vec{r}\}$, where $m\omega_0^2 \{\vec{r}\}$ is precisely the restoring force acting on the COM. When driven close to resonance, the plasmonic internal field is dramatically enhanced by a factor $\omega_0/\gamma = 2Q$ relative to the driving field, where $Q$ is the quality factor of the resonance. Since $Q$ can be large in high quality graphene disks, the total electric field can be dominated by internal fields with total electric fields well approximated by $e\vec E_{\rm tot} (t) \approx  m\omega_0^2 \{\vec{r}(t)\}$. 

Using Eq.~(\ref{eq:EtoA}) above and taking $e\vec E_{\rm tot} (t) \approx  m\omega_0^2 \{\vec{r}(t)\}$, we can rewrite $\tilde A_{L,R}$ in terms of the plasmon COM coordinates $\mathcal{Z}_\pm^{(0)}$. 
The conversion between the dimensionless (total) vector potential $\tilde A_{L,R}$ and the coordinate $\mathcal{Z}_\pm^{(0)}$ is given by the characteristic length $l = E_F \omega_d/(vm\omega_0^2)$ which naturally emerges from the definition of $\tilde{\vec{A}}_{L,R}$ in Eq.~(\ref{eq:fluxsupplement}), together with Eq.~(\ref{eq:EtoA}) and  $e\vec E_{\rm tot} (t) \approx  m\omega_0^2 \{\vec{r}(t)\}$. 
Re-writing $\tilde A_{L,R} \to \mathcal{Z}_\pm^{(0)}/l$ (with $L \to -$ and $R \to +$) in Eq.~(\ref{eq:fluxsupplement}) 
gives Eq.~(\ref{eq:FluxGen}) of the main text. 

\section{Continuous {\it vs}.~discontinuous transition}
As noted in the main text, the phase transition to the state with spontaneous non-equilibrium magnetization may occur either discontinuously as a function of driving amplitude (as in a first order phase transition), or continuously (akin to a second order phase transition).
The character of the transition follows from the $\eta$-dependence of the expression in brackets in Eq.~(\ref{eq:SelfConsistent}).
For fixed parameters, the zeros of this expression (as a function of $\eta$) determine the steady state values of magnetization that can be supported by the system.

As a first step in characterizing the nature of the transition, we note that $D_+D_-$ in Eq.~(\ref{eq:SelfConsistent}) is positive for all $\eta$; this must be so, as $D_+$ and $D_-$ are the squared absolute values of the denominators appearing in Eq.~(\ref{eq:PlasmonAmplitude}).
Consequently, for $\nu > 0$, nontrivial zeros of the term in brackets in Eq.~(\ref{eq:SelfConsistent}), and hence spontaneous magnetization, may only occur for $(\omega_d^2 + \gamma^2 - \omega_0^2) < 0$, i.e., 
to the left of the shaded region in Fig.~\ref{fig:LinearPolarization}{\bf b}.

The character of the transition (continuous {\it vs.}~discontinuous) is controlled by the sign of the quadratic term in $\eta$ in $D_+D_-$.
This can be seen by writing out 
$D_+D_- = A + B\eta^2 + C \eta^4$. 
Here $A$, $B$, and $C$ are coefficients that depend on damping and detuning, with $A, C > 0$:
\bea
\nonumber A &=& [(\omega_0^2 - \omega_d^2)^2 + \gamma^2\omega_d^2]^2\\
C &=& \nu^4 (\gamma^2 +\omega_d^2)^2.
\eea
The form of $B$ is straightforward to obtain from the explicit expressions for $D_+$ and $D_-$ given in the main text; the result is long and not illuminating, so we do not write it out in full here.
For $B>0$ (pink region, Fig.~\ref{fig:LinearPolarization}{\bf b}),  $(D_+D_-)^{-1}$ decreases monotonically as a function of $\eta^2$. 
When $E_{\rm rms}$ is small, the second term in the brackets in Eq.~(\ref{eq:SelfConsistent}) is less than one for all $\eta$, and no nontrivial solutions are found.
Above the critical value of $E_{\rm rms}$, a new branch of solutions emerges with $\eta$ growing continuously from zero. For $B < 0$ (blue region, Fig.~\ref{fig:LinearPolarization}{\bf b}), $(D_+D_-)^{-1}$ is non-monotonic, first increasing for small $\eta$, then decreasing for large $\eta$. Here, as $E_{\rm rms}$ is increased from zero, nontrivial solutions first appear at finite $\eta^2$ [corresponding to the maximum of $(D_+D_-)^{-1}$], and we obtain a discontinuous transition as in Fig.~\ref{fig:LinearPolarization}{\bf a}. 

\section{Self-consistent solutions of the steady state time evolution}
In the main text, we provide a detailed analysis of the steady states of the nonlinear dynamics described by Eq.~(\ref{eq:dynamicsfull}), under the approximation that the time-periodic Berry flux $\mathcal{F}(t)$ is replaced by its time-averaged (dc) part $\bar{\mathcal{F}}$. 
From a physical point of view, this approximation is motivated by the fact that it is the dc part of $\mathcal{F}$ that signifies a net breaking of TRS, and which we expect to be responsible for the instability towards a magnetized state.
From a technical point of view, this approximation introduces a vast simplification: by suppressing the time-dependent harmonics in $\mathcal{F}$, we ensure that Eq.~(\ref{eq:dynamicsfull}) supports solutions where $\vec{r}(t)$ and $\vec{p}(t)$ exhibit purely monochromatic oscillations (with no frequency components at higher harmonics of the driving frequency, $\omega_d$).
This simplification allows us to extract the time dependence $\sim e^{-i\omega_d t}$ from all variables, and solve a (nonlinear) algebraic equation for the steady state amplitudes of the left- and right- circulating components of $\vec{r}(t)$, $\mathcal{Z}^{(0)}_\pm$.
These solutions are given in Eq.~(\ref{eq:PlasmonAmplitude}).

To support our conclusions and to demonstrate that the time-dependent harmonics in $\mathcal{F}(t)$ do not significantly change the behavior of the system, we also performed self-consistent numerical simulations of the full equations of motion, Eq.~(\ref{eq:dynamicsfull}).
The simulations were performed as follows:
\begin{enumerate}
  \item We first initialize the system with values of position and momentum, $\vec{r}(0)$ and $\vec{p}(0)$, as well as an initial guess for the Berry flux $\mathcal{F}$.
The Berry flux $\mathcal{F}$ is specified by its dc part $\bar{\mathcal{F}}$ and a list of up to $n$ harmonics, $\mathcal{F}_1 \ldots \mathcal{F}_n$, such that $\mathcal{F}(t) = \bar{\mathcal{F}} + (\mathcal{F}_1e^{-i\omega_d t} + \ldots \mathcal{F}_ne^{-i  n\omega_d t} + {\rm c.c.})$. 
The value of $n$ is taken large enough to ensure convergence.

\item Next we numerically solve Eq.~(\ref{eq:dynamicsfull}) with the supplied form of $\mathcal{F}(t)$ for a large number of periods of the drive (typically of order 100), such that the system reaches a time-periodic steady state.

\item From the last several periods of evolution, we extract the left- and right-handed components of the motion via $\mathcal{Z}_\pm(t) = (x \pm iy)/\sqrt{2}$. From the Fourier transform of $\mathcal{Z}_\pm(t)$, we extract the values at its peaks centered around frequencies $\omega_d$, $2\omega_d$, etc.

\item We use the harmonics extracted from $\mathcal{Z}_\pm(t)$ to construct the time-periodic internal electric field and associated vector potential produced by the motion via $e\vec{E}_{\rm int} = m\omega_0^2\vec{r}$ and  $\vec{E}_{\rm int} = -\frac{1}{c}\frac{\partial}{\partial t}\vec{A}_{\rm int}$. 

\item \label{step:Berry} We numerically compute the Floquet band structure of the system using the total time-periodic field $\vec{A}_{\rm tot} = \vec{A}_{\rm drive} + \vec{A}_{\rm int}$.  We obtain a new time-periodic Berry flux $\mathcal{F}(t)$ by integrating the Berry connection $\mathcal{A}_{\vec{k}+}(t) = \Bra{\Phi_{\vec{k}+}(t)}i\nabla_{\vec{k}}\Ket{\Phi_{\vec{k}+}(t)}$ around the Fermi surface, $\mathcal{F}(t) = \oint d\vec{k} \cdot \mathcal{A}_{\vec{k}+}(t)$. Here $\Ket{\Phi_{\vec{k}+}(t)}$ is the time-periodic part of the Floquet state at crystal momentum $\vec{k}$ in the upper (+) Floquet band, derived from the original conduction band on the non-driven system.

\item Finally we return to step 1 and initialize the solver for Eq.~(\ref{eq:dynamicsfull}) with the final position and momentum of the previous iteration, and a new guess for $\mathcal{F}(t)$. This procedure is iterated until the Berry flux $\mathcal{F}(t)$ produced by the motion agrees with form that was used to compute it (i.e., until the change in $\mathcal{F}(t)$ from one iteration to the next falls below a convergence threshold).
\end{enumerate}

To improve stability of the code, we introduce an interpolating  
factor $\zeta$ such that the initial guess for the Berry flux for iteration $i$ + 1, $\mathcal{F}^{(i + 1)}$, is computed by interpolating between its value on iteration $i$ and the new value computed in step \ref{step:Berry} above: $\mathcal{F}_n^{(i+1)} = (1 - \zeta) \mathcal{F}_n^{(i)} + \zeta \mathcal{F}_n$.
Here $\mathcal{F}_n$ is the n-th harmonic extracted from the Fourier transform of $\mathcal{F}(t)$ computed in step \ref{step:Berry} above.
In the simulations shown we used $\zeta = 0.3$.
For each value of the driving amplitude we obtained convergence of all harmonics in $\mathcal{F}$ to better than 1 part in $10^6$.

We observe that, throughout the parameter regime studied, the higher harmonics of $\mathcal{F}$ decay very quickly with the order of the harmonic. (As a typical order of magnitude, we observe $\mathcal{F}_1/\bar{\mathcal{F}} \sim 10^{-3}$.)
Therefore we obtain rapid convergence with respect to the number of harmonics retained (in the simulations we keep the track the values of the first 5 harmonics of $\mathcal{F}$).

The procedure above was used to compute the green points shown in Fig.~\ref{fig:LinearPolarization}{\bf a}.
The good agreement with the solid curves confirms that the validity of our analytical treatment based on the dc part of $\mathcal{F}$.

\section{Estimate for critical field strengths}

In the following, we provide an estimate of the critical driving amplitudes required to achieve spontaneous out-of-equilibrium plasmonic magnetism. As an illustration, we will focus on the parameter regime where $B>0$ in Eq.~(\ref{eq:SelfConsistent}) of the main text. In this regime, the solutions of Eq.~(\ref{eq:SelfConsistent}) exhibit a continuous phase transition. In particular, non-trivial zeros of the bracketed expression in Eq.~(\ref{eq:SelfConsistent}) appear for $E_{\rm rms} > E_{\rm rms}^*$, where $E_{\rm rms}^*$ is the critical drive amplitude above which non-trivial solutions for $\eta$ appear:
\be
|e{E}_{\rm rms}^{*}|^2 = m^2\frac{\left[(\omega_0^2 -\omega_d^2)^2 + \gamma^2\omega_d^2\right]^2}{4\nu\omega_d(\omega_d^2 - \omega_0^2 + \gamma^2)} \approx \frac{m^2\omega_0^5}{4\nu}\cdot \frac{1}{4 Q^2}.
\ee
As a demonstration, in the second (approximate) equality we have set $\omega_d = \omega_0$, and used $\omega_0/\gamma = 2Q$ to express the result in terms of the quality factor. From the main text, recall that 
\be
\nu = \beta \kappa/l^2 = \beta \frac{m^3 v^2 \omega_0^6}{\hbar n_0 E_F^2 \omega_d^2}. 
\ee

Setting $\omega_d = \omega_0 = (2\pi e^2 n_0|\vec q|/m)^{1/2}$ and taking $|\vec q| = 1/d$, where $d$ is the diameter of the disk we estimate the critical field to be:
\be
|{E}_{\rm rms}^{*}| = 1.57 \times 10^2 \frac{(E_F [{\rm eV}]/0.1)^{7/4}}{\beta^{1/2} \times (Q/100) \times (d [\mu {\rm m}]/ 0.1)^{1/4}}[{\rm V}/{\rm cm}].
\ee
In obtaining the above estimate we have noted that, for graphene, the plasmon mass can be written as $m= E_F/v^2$, with $v= 10^8$ cm/s, and we have included flavor degeneracy for 4 valleys/spins when calculating the density. Recently, quality factors for plasmons exceeding $100$ have been realized, enabling even lower critical fields. 

The corresponding incident power required to induce the phase transition can be obtained by recalling (in SI units) $P_{\rm incident} = \epsilon_0 c|{E_{\rm rms}}|^2$, where $\epsilon_0$ is the permittivity of free space. Taking ${E}_{\rm rms}^* \sim 100\ {\rm V}/{\rm cm}$ as a rough order of magnitude estimate, we obtain a modest steady-state incident power of $P_{\rm incident}^* \sim 26 \, {\rm W} /{\rm cm}^{2}$. 

\section{Induced anisotropic Drude weight}

In the main text we focused on the nonlinearity that arises due to self-induced Berry flux.
In addition to the reconstruction of electronic wave functions that gives rise to the Berry flux, ac fields also modify the electronic dispersion.
In particular, a linearly polarized ac field, which breaks the rotational symmetry of the system, generically introduces anisotropy in the dispersion.
Such anisotropy in principle competes with spontaneous magnetization, as it favors a splitting of the plasmonic dipole resonance in the linear polarization basis (in contrast to magnetization, which is associated with circularly/elliptically polarized motion).
Below we estimate the anisotropy of the Drude weight (plasmon mass) induced by a linearly-polarized ac electric field.
We show that the induced anisotropy is small (of order $10^{-4}$) in the regime studied, and therefore should not qualitatively affect our results.

To estimate the induced anisotropy of the Drude weight, we consider a linearly polarized field $A_x \cos(\omega t)\, \hat{\vec{x}}= \tfrac{1}{2} A_x(e^{i\omega t} + e^{-i\omega t}) \hat{\vec{x}}$; note that for circularly polarized fields, no linear anisotropy develops. 
Using Eq.~(\ref{eq:graphene}) of the main text, and transforming to the ``extended space'' Fourier representation in which the Fourier harmonics $\{\ket{n}\}$ are treated as an auxiliary degree of freedom, 
the light-matter interaction is described via: $\mathcal{H}_{\rm int}  = \hat{V}\otimes \sum_n \left[  | n+1 \ra \la n|  + | n-1 \ra \la n|\right]$, where
\begin{align}
\hat{V}  = -\frac{E_F \tilde{A}_x}{2} \sigma_x. 
\end{align}
The eigenvalues of the Hamiltonian in the extended space representation are the quasienergies of the associated Floquet states.
Anisotropy in the Drude weight arises from EM field induced changes to the group velocity, different in the $x$ and $y$ directions, which in turn can be obtained from changes to the (quasi)-energy dispersion at the Fermi level. 

The lowest order EM-field induced changes to the energy occur at second order, so that $\varepsilon_{\vec k} = \varepsilon_{\vec k}^{(0)} + \varepsilon_{\vec k}^{(2)} $. 
Focusing on the case where the Fermi level is in the conduction band, the unperturbed energy is $\varepsilon_{\vec k}^{(0)} = v|\vec k|$, and the second-order correction is: 
\be
\varepsilon_{\vec k}^{(2)} = |\la \vec k, - |\hat{V}| \vec k, + \ra|^2\left(\frac{1}{2\varepsilon_{\vec k}^{(0)} + \hbar \omega} +  \frac{1}{2\varepsilon_{\vec k}^{(0)} - \hbar \omega}  \right),
\label{eq:secondorder}
\ee
where $| \vec k, \pm \ra$ are the unperturbed electronic states in the conduction and valence bands respectively, and $\hbar \omega $ is the energy of the absorbed/emitted photon. The first term in Eq.~(\ref{eq:secondorder}) arises from the virtual absorption of a photon accompanied by an electronic transition to a valence band state, followed by photon emission and a transition back to the conduction band; the second term occurs for a similar process, where the system emits a photon while making the (virtual) transition from the conduction to the valence band.  We note that there are two additional second order processes where photons are virtually absorbed/emitted without an accompanying electronic transition to the valence band. The corresponding terms in the energy shift have equal magnitude but opposite sign, and therefore cancel and do not contribute to the total $\varepsilon_{\vec k}^{(2)}$. 

Using $\la \vec k, - |\sigma_x | \vec k, + \ra = i \sin 2\phi$, where $\tan \phi = k_y/k_x$, we have:
\be
\varepsilon_{\vec k}^{(2)}  = \frac{E_F^2 |\tilde{A}_x|^2 \varepsilon_{\vec k}^{(0)}}{(2\varepsilon_{\vec k}^{(0)})^2 - (\hbar\omega)^2} \bigg[ 1 - \frac{1}{|\vec k|^2} (k_x^2 - k_y^2) \bigg].
\label{eq:secondorderresult}
\ee
Here we used the identity $\sin^2 2\phi = 1 - \cos^2 2\phi = 1 - (\cos^2 \phi - \sin^2 \phi)$. 
To find the change to the Drude weight, we use Eq.~(\ref{eq:secondorderresult}) to fit the perturbed energies $\varepsilon_{\vec{k}} = \varepsilon_{\vec{k}}^{(0)} + \varepsilon_{\vec k}^{(2)}$ to the anisotropic Dirac dispersion 
\begin{align}
\varepsilon_{\vec k} & = U + \hbar \sqrt{(v-\delta v)^2 k_x^2 + (v+\delta v)^2 k_y^2} \nonumber \\ 
\label{eq:anisoDirac}& = U + \varepsilon_{\vec k}^{(0)}- \varepsilon_{\vec k}^{(0)} \tfrac{ \delta v}{v|\vec k|^2} (k_x^2 - k_y^2) + \mathcal{O}(p^4),
\end{align}
where $U$ is a constant energy offset. Comparing the coefficient of the $k_x^2 - k_y^2$ term in Eq.~(\ref{eq:anisoDirac}) with Eq.~(\ref{eq:secondorderresult}) above, we obtain the velocity anisotropy as
\be
\frac{\delta v}{v} = \frac{E_F^2 |\tilde{A}_x|^2}{(2\varepsilon_{\vec k}^{(0)})^2 - (\hbar\omega)^2} \approx 4.4 \times 10^{-4}.
\label{eq:vanisotropy}
\ee
In the last (approximate) equality we have made an estimate using $\varepsilon_{\vec k}^{(0)} \approx  E_F = 160$ meV and $\hbar \omega = 100$ meV, as used for the figures in the main text, and have also used $|\tilde A_x| = ev\mathcal{E}_{\rm total}/(\omega E_F)  \approx 0.04$. 
For this last estimate, we used $\mathcal{E}_{\rm total} = 10^{4}$  V/cm for the total field (external drive + plasmonic internal) close to the phase transition, see section above. 
Note that the small value of $\tilde{A}$ that we find indicates that the changes/boost to the Fermi sea as the electron liquid moves is small as compared with the Fermi surface size. 
We also note that the $k$-independent term in Eq.~(\ref{eq:secondorderresult}) can be absorbed into the energy offset term $U$ with no changes to the velocity. 

Since the Drude weight $D_{ij} = e^2 \sum_{\vec k}  v_i (\vec k) v_j (\vec k) \delta(\epsilon_{\vec k} - E_F)$ scales as the square of the velocity at the Fermi surface, we estimate that the EM-induced anisotropy to the Drude weight (along $x$ or $y$ directions) is $ 2\delta v/v \approx 9 \times 10^{-4} $ for the regime described above in Eq.~(\ref{eq:vanisotropy}). 
Drude weight anisotropy can be directly included in the equations of motion in Eq.~(\ref{eq:dynamicsfull}) of the main text as a direction-dependent plasmon mass. 
We have numerically checked that incorporating anisotropy in the Drude weight of order $1\%$ does not change the qualitative picture of our results, and only introduces modest quantitative changes to the threshold driving amplitude and induced Berry flux. As a result, given that the induced anisotropy is expected to be an order of magnitude smaller, we conclude that EM-induced anisotropies do not adversely affect the realization of the non-equilibrium TRS breaking phase transition we discuss in the main text.

\end{document}